\documentclass[12pt]{iopart}
\usepackage{graphicx}
\usepackage[vflt]{floatflt}
\begin{document}

\title[Light Vector Mesons from dAu in PHENIX]{Light Vector Mesons from dAu in PHENIX}

\author{Richard Seto\dag\ for the PHENIX Collaboration\ddag}

\address{\dag\ Physics Department, University of California, Riverside California, 92521 USA}
\address{\ddag\ For the full PHENIX Collaboration authors and acknowledgments, see ``Collaborations'' appendix of this volume.}

\begin{abstract}
A first measurement of the $e^+e^-$ decay rate of $\phi$ mesons in
dAu collisions from the PHENIX detector at RHIC and its comparison to
the $K^+K^-$ decay channel is described. The comparison of the two 
decay channels can be sensitive to chiral symmetry restoration.
\end{abstract}



\section{Introduction}
Deconfinement is the first and most familiar manifestation of the 
QCD phase transition. Of equal importance is the chiral
transition. The spontaneous breakdown of chiral
symmetry responsible for the masses of the hadrons.  The study of
low mass vector mesons is an ideal probe of chiral
symmetry restoration in heavy ion collisions.  As an example, the 
lifetime of the $\phi$ is about 50 fm/c in the vacuum and a fraction 
of $\phi$'s
produced in a heavy collision will decay in the fireball.
Decays to di-leptons are particularly attractive since the daughter
particles are not strongly interacting and can reach the experimental
apparatus without rescattering, hence any effects such as mass shifts
or broadening of spectral functions due to the onset of chiral
symmetry restoration would show themselves in the invariant mass
spectra. The E325 experiment at KEK\cite{E325} and CERES at CERN
\cite{CERES} have seen hints of such effects in both pA and AA
collisions.

Because of the small Q value of the KK decay channel, a comparison of
the $\phi \rightarrow$ KK/$\phi \rightarrow$ ee rate is a particularly
sensitive measure of a mass modification of either the $\phi$ or the
kaon~\cite{KKEE}. The PHENIX detector is ideally suited for such
measurements in that it has good particle identification for kaons
with momentum between 300 MeV/c and 2 GeV/c and electrons to 5 GeV/c,
as well as good momentum resolution. All data presented in this paper
are from deuteron-Gold collisions at $\sqrt{s_{NN}}$=200 GeV measured
by PHENIX in 2003.

\section{di-electron Analysis}
About 31M single-electron triggered events were used for the electron
analysis. The trigger required matching hits in the Ring Imaging
Cerenkov Counter(RICH) and the Electromagnetic Calorimeter (EMC),
where the threshold was set to 600 MeV. An additional data set with a
higher threshold was not used for the present analysis.  Electrons
were identified by requiring two or more phototubes firing in the
RICH and an E/p match between the energy measured by the EMC and the
momentum measured by the tracking system. Specifically we required
$0.5<E/p<1.5$.  Cuts are then made to remove electron candidates
coming from conversions. The invariant mass spectrum of $e^+e^-$ pairs
is then formed as shown in figure 1.

In order to form a background sample, opposite sign pairs were taken
from different events. Because our sample was triggered, care was
taken that the mixed event background had the same characteristics as
the events themselves.  In particular minimum bias events were used to
form the mixed pair, where one of the electrons was required to pass
the trigger requirements. Additionally, event used to form the mixed pair
 were required to have similar
centralities and vertex positions.  Normalization of the background
distribution was done by matching the data to the background in a
sideband region between 850 and 950 MeV below the signal and 1100-1200
MeV above the signal. Other methods of normalization were used to
estimate the systematic error from this procedure.

\begin{figure}
\centering
\begin{tabular}{cc}
\begin{minipage}{3in}
\includegraphics[width=3in]{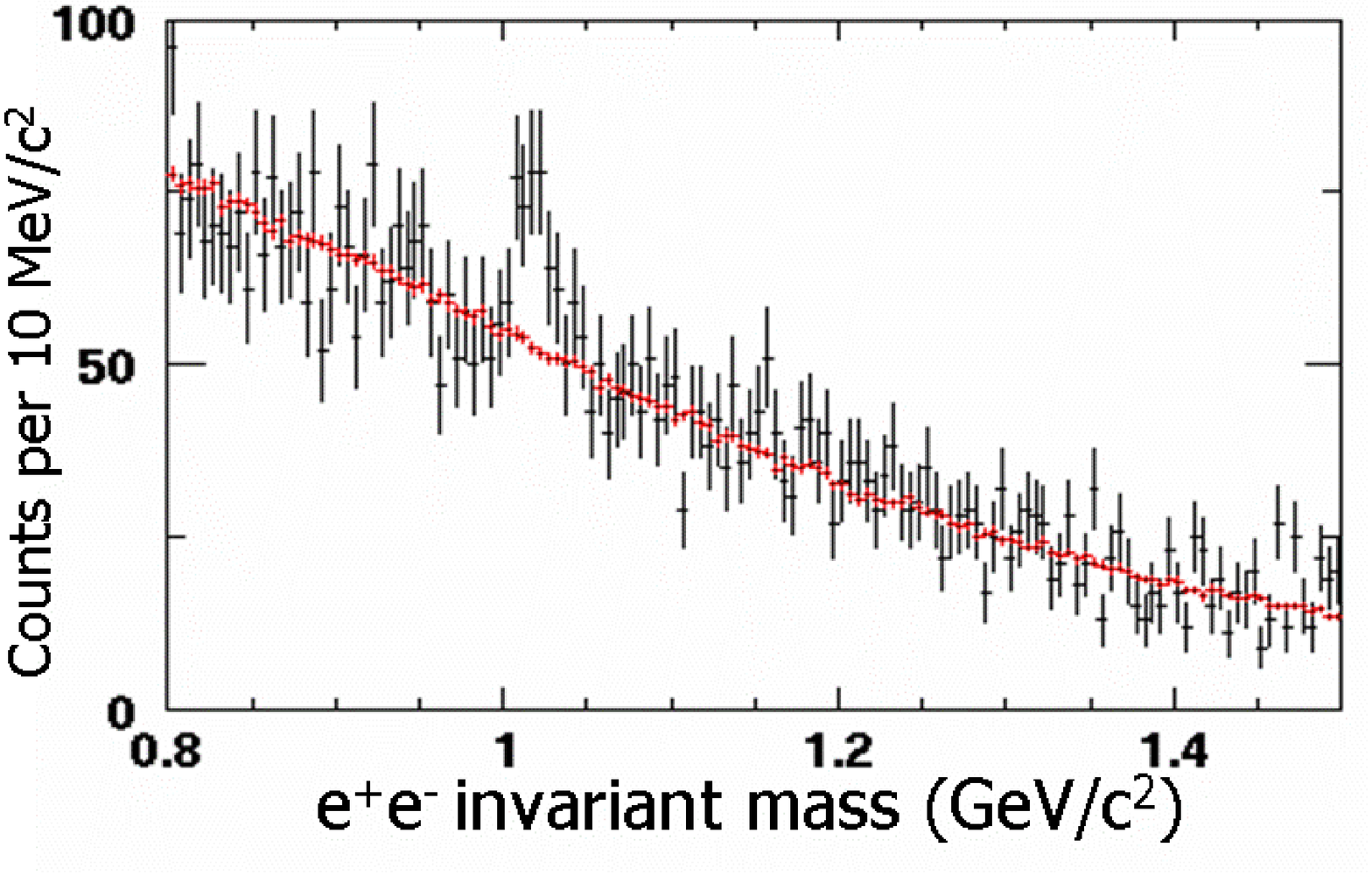}\caption{$e^+e^-$ invariant mass for minimum bias events, with the background superimposed. } 
\end{minipage}
\begin{minipage}{3in}
\includegraphics[width=3in]{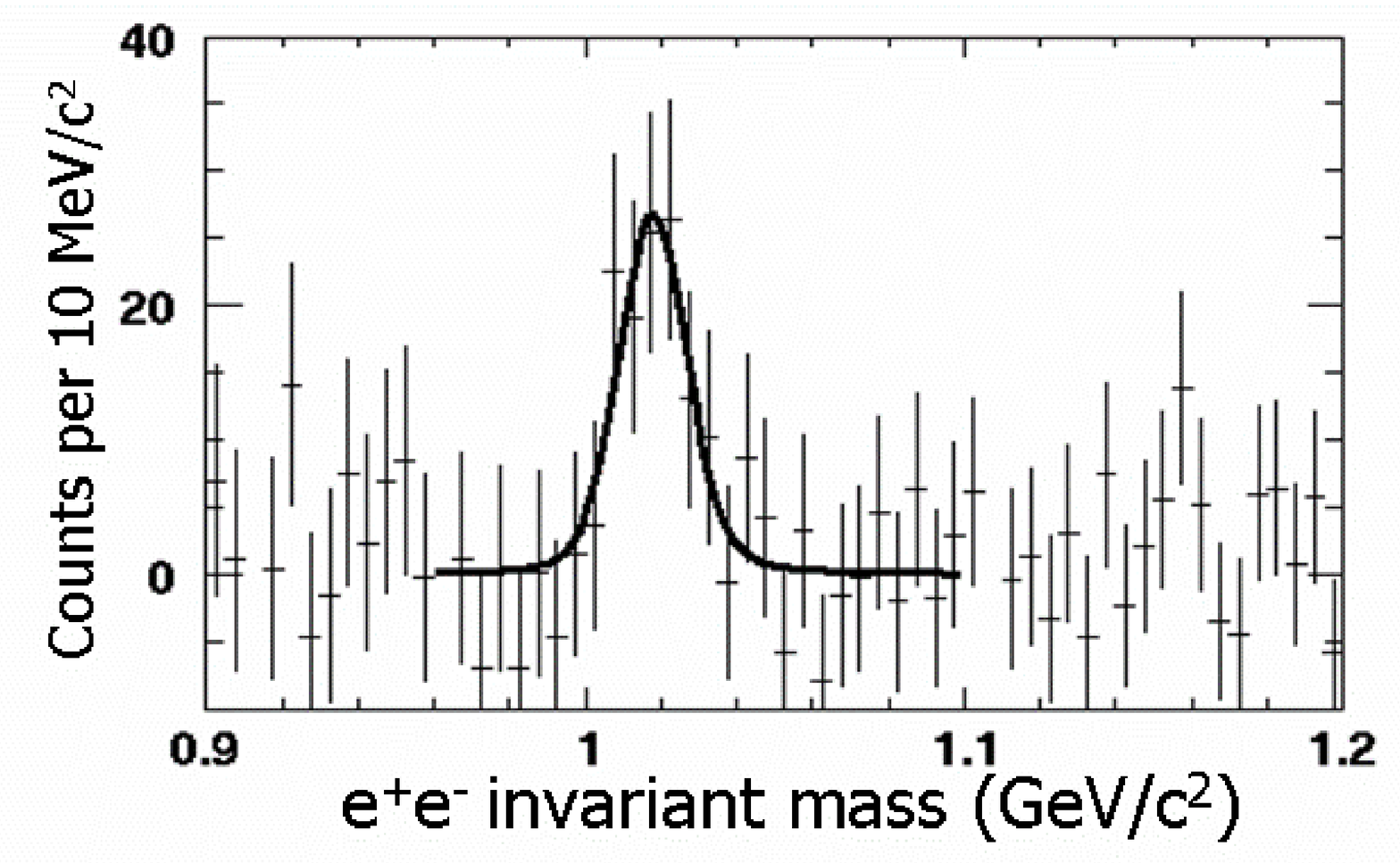}\caption{Background subtracted $e^+e^-$ invariant mass spectrum with a fit as described in the text.} 
\end{minipage}
\end{tabular}
\end{figure} 

The background subtracted signal was then fit to a relativistic
Breit-Wigner convoluted with a Gaussian to account for the
experimental resolution (figure 2). The width $\Gamma$ was held fixed at the
particle data book value. Values obtained from the fit are shown in
Table~\ref{tab:tc}.  The value of the mass is consistent with the
known value of the mass in the vacuum. In addition the fitted
experimental resolution is consistent with simulations of the detector
performance. The signal was then divided into 3 bins of $M_T$, and
corrections were made for acceptance and efficiencies. A fit was then
done to the invariant yield resulting in the yield and inverse slope
shown in the table. Major contributors to the systematic error on
dN/dy are the normalization of the background and its effect on the
inverse slope, and the run-by-run variations from the
electron-trigger.
\section{$K^+K^-$ Analysis}
\begin{floatingfigure}[r]{3.4in}
\includegraphics[width=3.2in]{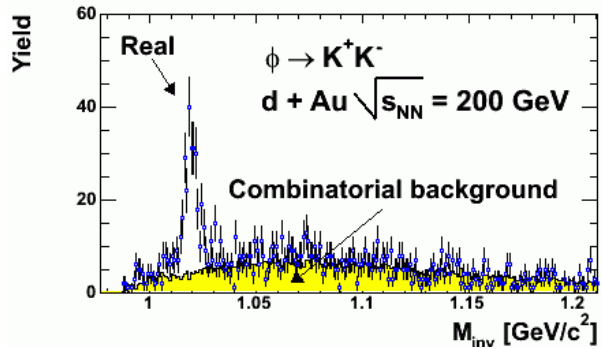}\caption{$K^+K^-$ invariant mass spectrum with the background superposed.} 
\end{floatingfigure} 
About 62M minimum bias events were used for the $K^+K^-$
analysis. Kaons were identified in the time-of-flight detector (TOF)
which covers $\Delta\phi\approx 40^\circ$. Because of
the limited coverage of the TOF, the analysis preferentially accepted
higher momentum $\phi's$ as compared with the electron analysis. The
invariant mass spectrum was generated by combining unlike sign pairs of
kaons. The background shape was formed in a similar manner to the
electrons by mixing pairs from events with similar centralities and
vertices. The normalization in this case was determined by adjusting
the integral of the background to $2\sqrt{N_{++}N_{--}}$ where
$N_{++}$ and $N_{--}$ refer to the like sign combinations in each
event. Once again the signal was fit to a relativistic Breit-Wigner
convoluted with a Gaussian for the experimental resolution. Since
the resolution of the $\phi$ in the KK decay channel is rather
insensitive to momentum effects, we chose to hold the experimental
resolution fixed to the value obtained from the simulation - 1.2
MeV. Results of the fit are shown in the table and figure 3. Both the mass and
width are consistent with PDG vacuum values. For both decay channels,
$\Gamma$ and $\sigma_{exp}$ are highly correlated, when they are both
allowed to vary in the fit.

The data are then divided into bins of $M_T$, corrections are made for
acceptance and efficiencies and a fit is done to obtain the yield.
Because of the good signal to background in the KK channel, systematic
errors are considerably smaller than in the electron channel. The
systematic errors are dominated by the range in $M_T$ over which the
fit was done, the run-by-run changes in efficiency, and the
corrections due to the fiducial cuts.  Figure 4 shows a comparison of
the data points from both the electron and kaon analyses. The fit
shown in the figure is to all of the points. When fits were
done to extract yields for comparison, the fits were done separately
for the electron and kaon channels. The two data samples are
consistent in both the yield and the inverse slope as can be seen by
looking at the comparisons in figures 5 and 6 of the extracted inverse
slopes and yields.
\vskip .1in
\section{Conclusions}
Within the substantial errors - particularly from the electron analysis -
the yields of the $\phi\rightarrow ee$ and $\phi \rightarrow KK$ are
consistent with one another. Because of the small fraction which decay
inside the relevant volume, the effects from chiral symmetry
restoration, particularly in dAu collisions, are expected to be
small. PHENIX will make further improvements to this analysis by
including the data with a higher trigger threshold and improving
systematic errors - particularly from background normalization in the
electron channel.  PHENIX also sees a clear $\omega \rightarrow ee$
signal, which, with a shorter lifetime than the $\phi$ will increase
the number of decays in the fireball.  In addition, effects in Au-Au
collisions could be considerably stronger since energy densities are
higher. Further in the future, upgrades to PHENIX such as the Hadron
Blind Detector (HBD) will significantly enhance our capabilities to
reject conversion and Dalitz pairs which are the dominant source of
background. This will open up the possibility of studying the $\rho
\rightarrow ee$ as well as thermal di-electrons.
\begin{figure}
\centering
\begin{tabular}{cc}
\begin{minipage}{2.6in}
\includegraphics[width=2.6in]{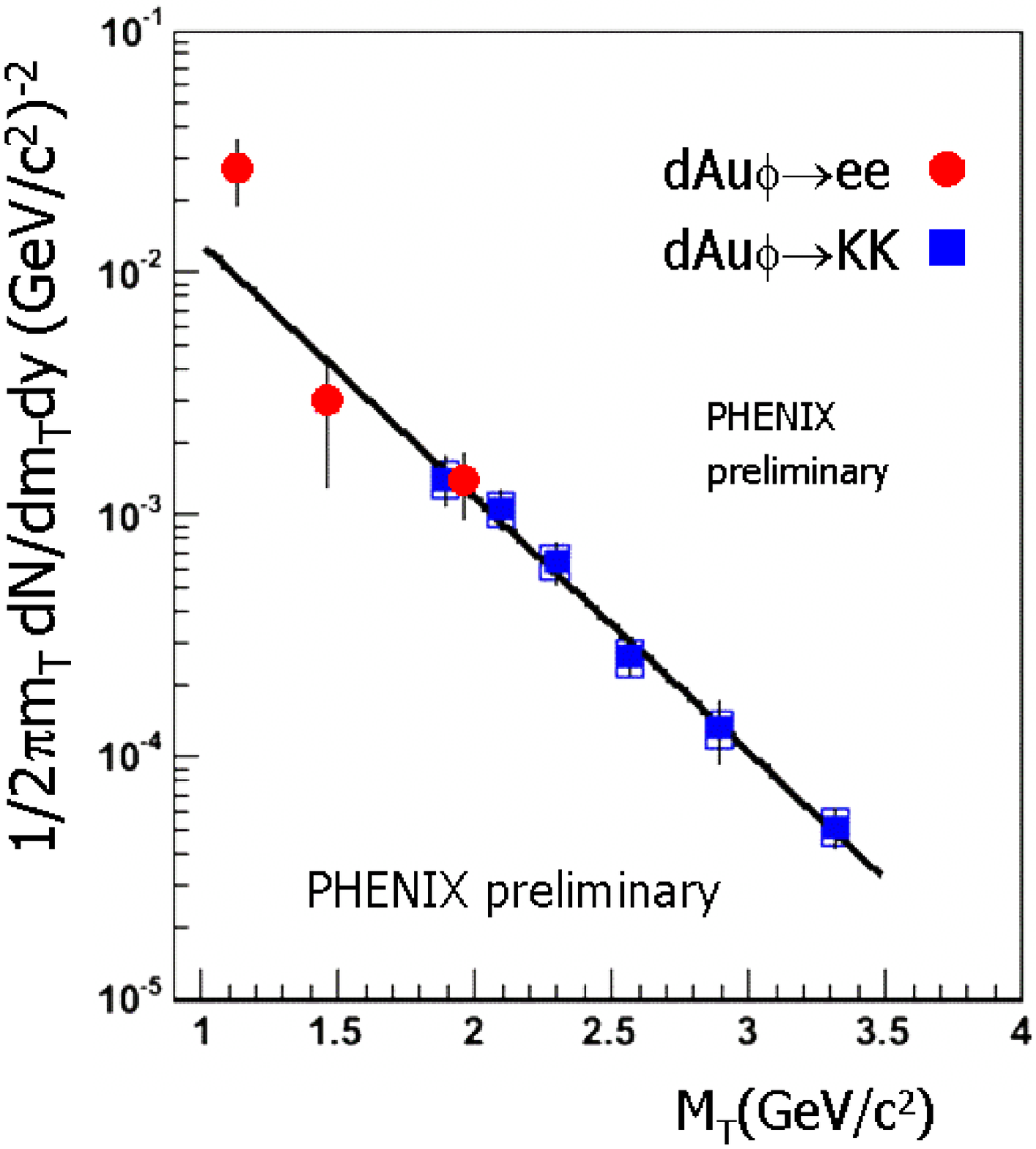}
\caption{$M_T$ spectrum for minimum bias events from both the $e^+e^-$ and $K^+K^-$ channels. The fit is to all the points.} 
\end{minipage}
\begin{minipage}{3.2in}
\includegraphics[width=3.2in]{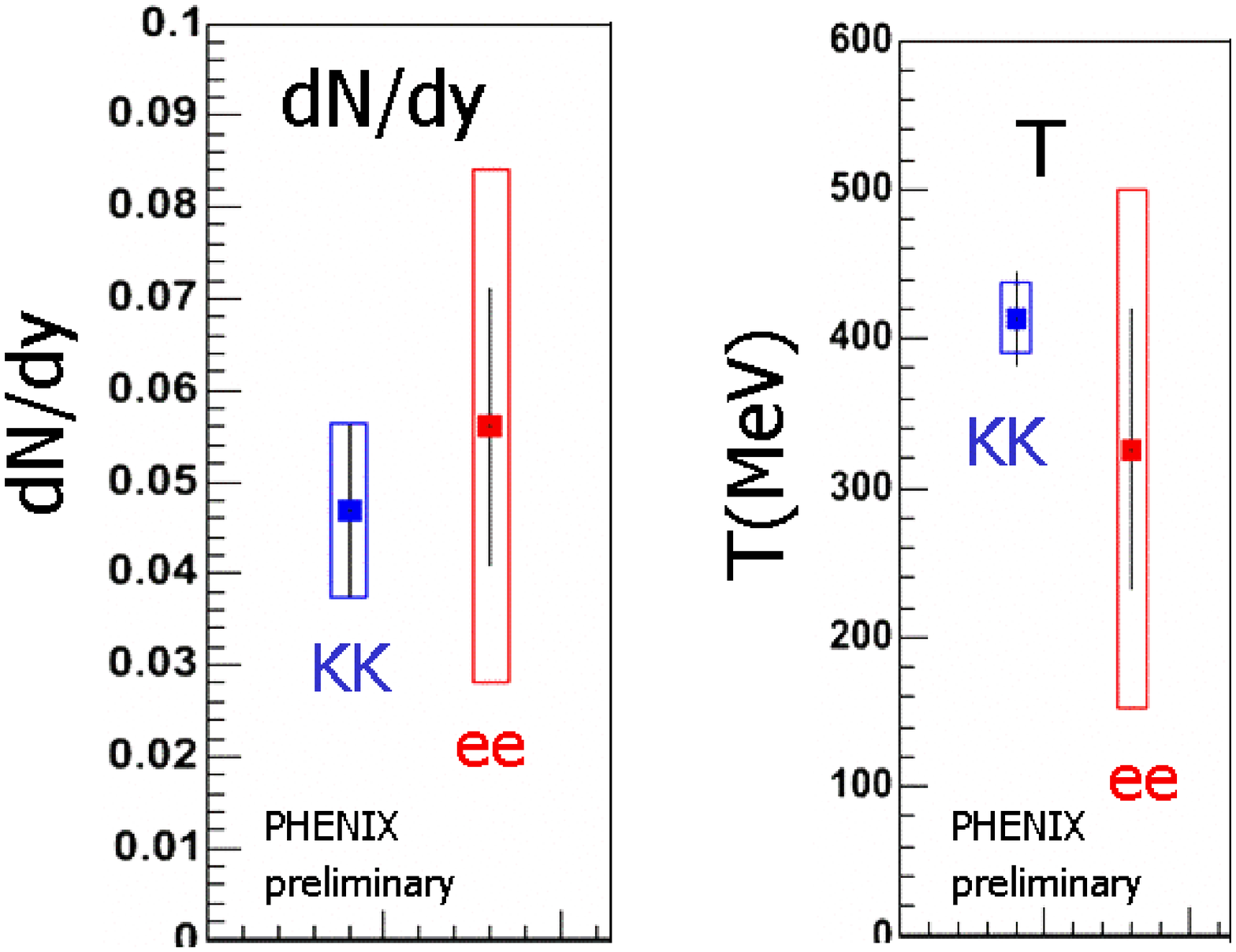}
\caption{A comparison of the inverse slope(right) and yields(left) of the $\phi$ } 
\end{minipage}
\end{tabular}
\end{figure} 
 
\begin{table}[h]
\begin{center}
\begin{tabular}{|l|l|l|l|}
\hline
                        &$\phi\rightarrow ee$& $\phi\rightarrow KK$& PDG\\
\hline
Mass $(GeV/c^2)$                    &1.0177$\pm$0.0023 &1.0193$\pm$0.0003 & 1.01946\\
\hline
$\Gamma$ $(MeV/c^2)$               & 4.26(fixed)& 4.750$\pm$0.67 & 4.26 \\
\hline
$\sigma_{exp} (MeV/c^2)$ & 8.1$\pm$2.1& 1.2(fixed) & \\
\hline
N &120 & 207 & \\
\hline
dN/dy & 0.056$\pm$0.015$\pm$50\% & 0.0468$\pm$0.0092$^{+0.0095}_{-0.0092}$& \\
\hline
T (MeV)&326$\pm$94$\pm$53\%& 414$\pm$13$\pm$23& \\
\hline
\end{tabular}
\caption {\label{tab:tc}Results of fits to the invariant mass spectra and
the yields for both the electron and kaon decay channels of the $\phi$. Statistical errors are listed first, followed by systematic errors.}
\end{center}
\end{table}


\begin{thebibliography}{99}

\bibitem{E325}  K.Ozawa et al, Phys. Rev. Lett. 86-22 5019-5022(2001) 

\bibitem{CERES}  CERES Collaboration, G. Agakichiev et al, Phys. Lett. B422 (1998) 405 

\bibitem{KKEE}          D. Lissauer and E. V. Shuryak, Phys. Lett. B253, 15 (1991). 

\end{thebibliography}
\end{document}